# Power Grid Cybersecurity: Policy Analysis White Paper

**Jack Vanlyssel**
**GLNS 500**
**10/8/2024**

# Executive Summary

The U.S. power grid is critical to national security, public safety, and economic stability. It provides foundational infrastructure that powers essential services, such as defense operations, emergency response systems, and commercial activities. With vulnerabilities in industrial control systems, network protocols, power grid devices, remote access, supply chains, poor cyber hygiene, and a lack of a cyber recovery plan, the power grid is a high-value target at risk of attack. Despite the escalating threat, the U.S. lacks effective policy that could protect its most critical infrastructure.

To remedy the issue, a dual policy approach emphasizing enhanced information sharing and standardized cyber hygiene practices would immediately improve power grid security. Secure information-sharing platforms between the government and private sector will allow faster threat detection and response and reduce vulnerability through shared intelligence. Mandatory cyber hygiene standards would prevent hackers from using common exploits, such as weak passwords and phishing emails, to compromise systems. These two policies will efficiently strengthen power grid cybersecurity in the short term.

To ensure long-term security, the U.S. needs a Unified National Cybersecurity Framework that consolidates existing cybersecurity standards from NERC, IEC, IEEE, and NIST. This framework would reduce regulatory overlaps, ensure consistent security protocols nationwide, and be adaptable to evolving cyber threats. This policy would require more time and significant coordination between federal agencies, state regulators, and private utilities, but it would immensely benefit U.S. power grid security.

The U.S. power grid is essential for national security and economic stability. Although no attacks have been recorded out as of now, the threat to our power grid has never been greater. By enacting the policies recommended in this paper, the U.S. can better defend its critical infrastructure against evolving cyber threats and ensure the protection of essential services for years to come.

# Table of Contents

# List of Acronyms and Abbreviations

CRISP - Cybersecurity Risk Information Sharing Program

DHS - Department of Homeland Security

DOE - Department of Energy

E-ISAC - Electricity Information Sharing and Analysis Center

GAO - Government Accountability Office

ICS - Industrial Control Systems

IEC - International Electrotechnical Commission

IEEE - Institute of Electrical and Electronics Engineers

IT - Information Technology

NERC - North American Electric Reliability Corporation

NIST - National Institute of Standards and Technology

OT - Operational Technology

SCADA - Supervisory Control and Data Acquisition

# Introduction

## Power Grid Fundamentals

The United States power grid is split between zones: the Western Interconnection, the Eastern Interconnection, and the Texas Interconnection. The Western and Eastern Interconnections divide the country along the borders of New Mexico, Colorado, Wyoming, and Montana, while the Texas Interconnection covers most of Texas and operates independently. These power grids comprise three key components: generation, transmission, and distribution. Generation involves power plants that produce electricity from sources like coal, natural gas, wind, and nuclear power, often in remote locations. Transmission consists of high-voltage lines that transport electricity over long distances from generators to areas closer to consumers. Finally, distribution comprises the power infrastructure that provides power to homes and businesses. Substations are located between generation, transmission, and distribution and contain transformers that increase electricity voltage for transport and decrease electricity voltage for public distribution [1].

## Governance of the U.S. Power Grid

The three interconnections that make up the power grid are themselves made up of numerous entities. Private utility companies, independent power producers, electrical cooperatives, and government owned facilities all work together to operate the grid on a daily basis. This decentralized ownership creates a variety of generation, transmission, and distribution assets, each with unique standards, protocols, and operational practices [1]. Coordinating cybersecurity across these stakeholders is challenging because it requires effective collaboration and trust between the public and private sectors. Unlike a centrally controlled system, the U.S. power grid must navigate many different interests, regulatory requirements, and operational priorities, making unified cybersecurity measures far more complex. In addition, private companies may need more direct incentives to invest in grid security as their legal liability for grid failures or cyberattacks can be unclear or limited. The complex nature of grid ownership has been a major contributing factor to cybersecurity vulnerabilities in America.

## Smart Grid

As aging parts in the grid are replaced with new components, utilities in the U.S. are increasingly incorporating smart grid technology with inherent cyber vulnerabilities into their operations. Smart grids use digital technology to manage electricity flows, allowing for real-time data exchange between consumers and utilities to optimize energy usage. This shift transforms the grid from a disorganized system of power producers and consumers into an interconnected network that could be brought down by coordinated cyberattacks [2]. While no reported cyberattacks have resulted in long-term damage to U.S. power system operations, emerging risks leave the power grid vulnerable to attacks that threaten reliability and potentially lead to large-scale power outages.

## Problem Statement

Electricity is the backbone of American society, powering everything from homes and businesses to our national defense systems. As the grid has become more interconnected, power companies and utilities around the world have seen a six-fold increase in the number of detected cyber incidents [3], yet the U.S. lacks a comprehensive policy approach for securing its energy infrastructure. Vulnerabilities in Industrial Control Systems, network protocols, power grid devices, remote access, supply chains, and poor cyber hygiene, coupled with a lack of a cyber recovery plan, demand government intervention. Without action, there will be catastrophic consequences for U.S. national security, public safety, and economic stability.

# Current Power Grid Vulnerabilities

## Insufficient Network Protocols

Utilities rely on networks to connect equipment and systems, but these networks often have vulnerabilities. A 2014 survey found that insecure IT networks accounted for 41% of security incidents at electric utilities [4]. Weak security protocols like Modbus and DNP3 are widely used in power grid network communications across the country. Both lack encryption and authentication, allowing attackers to easily intercept data as it is being transmitted. Network misconfiguration also exposes critical systems to unauthorized access, allowing attackers to bypass security controls and potentially disrupt or control key infrastructure operations. These issues stem from most utilities' lack of proper network administration, which indicates a need for refined standards and new protocols.

## Industrial Control Systems

In the power grid, Industrial Control Systems (ICS) monitor and control critical operations across generation, transmission, and distribution stages. Traditionally, ICS have run on Operational Technology (OT) networks completely separate from the Information Technology (IT) networks that are used to run the broader facility housing the ICS. With the adoption of smart grid technologies, OT networks are being integrated with IT networks. This enables attackers to exploit vulnerabilities in internet-connected IT systems to gain access to ICS, which used to be wholly disconnected from the internet. The previous reliance on network segmentation to protect ICS has left control systems of the power grid open to attack. The generation sector is particularly vulnerable, as a compromised ICS can cause physical damage and potentially disable generators, as demonstrated by a 2007 Idaho National Lab experiment that physically destroyed a diesel generator manipulating protective relays accessible through IT systems [5].

## Power Grid Devices

Many power grid components, such as transformers and circuit breakers, were designed with a focus on operational reliability. They often lack essential security features like encryption and authentication and are incompatible with modern security protocols. Similar to ICS, these specialized devices are being connected to modern IT networks. Many utilities don't consider the security issues associated with connecting these devices, which creates many security risks. To make matters worse, attackers can use search engines like SHODAN to easily locate these connected components openly on the internet [6].

## Remote Access & Mobile Devices

Utilities use remote access and mobile devices to manage geographically widespread assets, but security measures such as strong passwords and encryption are often neglected. In a 2014 DHS report, a public utility (whose name was not disclosed) was compromised by nation-state-backed hackers through remote access due to weak security configurations [7]. This compromise occurred because the utility had inadequate protections on its remote access system. Utilities will continue to expand the use of remote access technology into the future, and if left unsecured, similar incidents will occur.

## Third-Party Services and Supply Chains

Smart grid devices often rely on components from international manufacturers, which could have built-in exploits, creating supply chain vulnerabilities in many power grid networks. Third-party service providers also present risks. Vendors can be slow to patch security flaws or miss them entirely, leading to vulnerability. Power grid operation relies on software from companies like Microsoft, Siemens, and SolarWinds, which have all been exploited in recent high-profile cyberattacks [8].

## Lack of Cyber Hygiene

Cyber hygiene refers to procedures to maintain the security of devices, systems, and data in a digital environment. Many utilities are lacking in these practices. For example, weak or default passwords are used, making it easy for attackers to gain unauthorized access. Organizations don't regularly update their software. Data backups are often ignored. Worst of all, employees fall victim to social engineering attacks such as phishing or fake phone calls that are used to steal passwords or sensitive information, which is one of the most common entry methods attackers use to access systems [9].

## No Cyber Recovery Plan

The U.S. has experience with power grid recovery from natural disasters such as Hurricane Sandy but lacks a strategy for recovering from a power grid cyberattack. If the power grid were shut down, as in the 2015 attacks on Ukraine, the U.S. might not be able to restore power efficiently or at all in the worst case [9]. The lack of preparedness may result in increased damage to infrastructure, economic losses, and delays in restoring power, further compromising national security [10].

# Root Causes of Vulnerability

## Fragmented Policy and Standards

The fragmented policy landscape in the U.S. energy sector undermines power grid cybersecurity efforts. Organizations, including NERC, IEEE, and NIST, have created distinct cybersecurity standards for utilities to follow [11]. In addition to national standards, states also have unique security standards that grid operators must adhere to, which vary widely based on local priorities and resources. On top of this, different sectors of the grid, such as nuclear or renewable energy, have different sets of standards and regulations. NERC is the only organization that has the authority to enforce cybersecurity standards at a national level, and due to the overwhelming amount of clashing rules and regulations, it's challenging to enforce any standards at all. This issue will prevent the power grid as a whole from ever achieving a baseline level of cybersecurity.

## Lack of Risk Assessment Framework and Legal Definitions

The lack of clear legal definitions for security concepts adds to the challenge of fragmented standards. NIST, for example, defines "cybersecurity" as protecting information systems from unauthorized access, disruption, or destruction, while IEEE focuses on safeguarding critical infrastructure, like power grids, by securing communication protocols and operational technologies [12]. These varying perspectives lead to differing priorities in security measures depending on how the language in the policy is interpreted. Additionally, there is no system in place to assess or quantify cyber vulnerabilities. Organizations who can't afford full time cybersecurity specialists are left without a way of knowing what vulnerabilities they have, leaving large security gaps throughout the power grid. Without the ability to talk about and classify security issues, utilities will remain at risk.

## Insufficient Public-Private Partnerships

Government agencies and national laboratories have the most up to date and impactful information regarding cyberattack strategies used against utilities and grid operators. The government does not adequately share this knowledge throughout the energy sector, where it can be deployed by utilities to defend their infrastructure. Government programs like the Cybersecurity Risk Information Sharing Program (CRISP), which facilitates real-time sharing of threat data between utilities and the government, attempt to solve this issue but remain ineffective. Companies are resistant to use CRISP and similar programs due to concerns over potential liability from sharing information and a lack of perceived benefits [13]. The Electricity Information Sharing and Analysis Center (E-ISAC), which provides security information and analysis for the energy sector, has not done enough to demonstrate the value of information sharing, especially for smaller companies, to enhance participation [13].

## Limited Resources, Training, and Human Error

Many utilities lack security personnel, and cybersecurity decisions are often made at the executive level without input from professional staff, creating vulnerability. In a survey from the Foundation for Defense of Democracies, sixty percent of utilities spent less than five percent of their budget on IT security in 2021, and the smallest utilities spent less than one percent on security [14]. Without money to hire trained staff, utilities will often only employ the most basic cybersecurity protocols. Grants such as the Energy Independence and Security Act provide funding for utilities to invest in cybersecurity, but with 70 percent of utilities reportedly employing less than three full time security personnel, more resources are needed to safeguard critical infrastructure effectively [14].

# Stakeholder Impact

## Primary Stakeholders

A successful cyberattack on the U.S. power grid would impact utilities, government entities, and the general public. Utilities would face operational, financial, and reputational setbacks, including service disruptions, equipment damage, and legal consequences. Federal and state agencies, such as the Department of Energy and the Department of Homeland Security, would have to reevaluate cybersecurity standards, coordinate the response to the attack, and ensure public safety. The general public would suffer from power outages and disruptions to essential services like healthcare, transportation, and emergency response systems, leading to substantial economic consequences. Power disturbances already cost at least $119 billion annually, and single events like the 2003 blackout have cost up to $6 billion [15].

## Secondary stakeholders

Secondary stakeholders, including the national security apparatus, private sector businesses, and critical infrastructure operators, are also significantly impacted by cyber threats to the power grid. The national security apparatus, which relies on a secure power supply for communications, operations, and logistics, would face severe compromise from any disruption to the grid. A 2009 Government Accountability Office report found that 31 of the Department of Defense's 34 most critical assets rely on commercial power, with 99% of DoD energy originating from private companies with insufficient backup capacity for prolonged outages [16]. Private sector businesses in finance, healthcare, manufacturing, and transportation would also experience significant financial losses and reputational damage from power disruptions, while other critical infrastructure sectors, such as water supply and communications, risk cascading failures that could bring all of America to a standstill. Defense Secretary Leon Panetta described a cyberattack on the power grid as a "cyber Pearl Harbor" to emphasize the severe national security implications that would be felt throughout the entire country.

# Policy Analysis

## Preface

Due to the variety of vulnerabilities that have been discussed, a range of policies will be offered for remediation. Each policy will be evaluated across four criteria and assigned a rating of Low, Medium, or High in order to identify the most effective option. The criteria used are overall impact on security, feasibility, adaptability, and implementation cost. After all policies have been assessed, a policy matrix will be used to recommend the most effective policy for implementation based on an average of how well policies perform in each category. Please note that implementation costs will be rated inversely to the other options, with low costs being favorable.

## **Unified National Cybersecurity Framework**

Overall Impact on Security - **High**

Establishing a unified national cybersecurity framework would significantly enhance the security of the U.S. power grid. Consolidating cybersecurity standards from organizations such as NERC, IEC, IEEE, and NIST into one framework would reduce confusion and overlap from existing policy. The framework would also allow the government to establish clear definitions for concepts like the smart grid or cyberattacks, giving utilities the legal clarity needed to align their defense strategies. A single, cohesive framework would help ensure that all utilities, regardless of size or location, adhere to a baseline of cybersecurity measures, thus reducing gaps in security across the board [17]. Without this policy, implementing a cybersecurity baseline will continue to be a challenge.

Implementation Cost - **High**

Implementing a unified framework would be costly. The government would need to set aside a large amount of capital to fund the research, development, and implementation of a new power grid cybersecurity framework. The framework also requires extensive workforce training and infrastructure upgrades for utilities with outdated systems. In addition to the monetary cost, the creation and implementation of this unified framework could potentially take years. Disaster mitigation and more efficient standards may offset upfront costs in the long term, but the initial buy-in remains a significant challenge [17].

Feasibility - **Medium**

While the policy has a high potential impact, coordinating multiple federal and state agencies is a complex task. To create the framework, the government would need to choose an agency to lead the project and engage with stakeholders across the power grid in order to develop a plan that would not be rejected by utilities. Implementing the framework would be less challenging because an agency such as NERC could effectively enforce unified standards where, as before, the overwhelming amount of standards made enforcement impossible [18]. Federal leadership and phased implementation would help improve feasibility, but the challenges of aligning numerous public and private stakeholders must be considered.

Adaptability - **High**

A unified framework would offer significant adaptability, allowing the government to update the entire power grid's baseline defenses as new cyber threats emerge and technology advances. This would ensure that cybersecurity practices, particularly in smart grid and ICS protection, evolve into the future [17].

## Cyber Hygiene Standards

Overall Impact on Security - **Medium**

Energy companies identify phishing as their largest vulnerability today [19]. Creating cyber hygiene standards for utilities, such as social engineering awareness training, phishing training, and strong password requirements, would immediately raise the baseline level of security across the grid. While these standards would protect the grid from common vulnerabilities like phishing and malware, their overall effect is moderate, as they primarily focus on strengthening basic security measures. More advanced defenses would still be necessary to protect against more sophisticated cyberattacks.

Implementation Cost - **Low**

The cost of implementing cyber hygiene standards is comparatively low. Most expenses would stem from personnel training and deployment of basic network monitoring tools. Many utilities already have basic cyber hygiene, so this policy would primarily standardize and enforce these practices across the industry. Overall, the cost is modest compared to more comprehensive policies [20].

Feasibility - **High**

Many cyber hygiene methods are already recognized as best practices within the industry. For example, the NIST Cybersecurity Framework, which is widely used in the energy sector, recommends implementing strong password policies, regular system updates, and training employees to recognize phishing attempts as protective measures [21]. Additionally, NERC has incorporated some cyber hygiene principles into its Critical Infrastructure Protection standards, making essential cybersecurity practices necessary for utilities. Formalizing these measures would be relatively easy to implement across the sector because cyber hygiene primarily involves personnel training and minor system upgrades, requiring less coordination and fewer resources than other cybersecurity policies.

Adaptability - **Medium**

Cyber hygiene standards can be updated to address evolving threats, so it is an adaptable policy. Its weakness is that more sophisticated attacks require defenses beyond basic hygiene practices, so the policy is only adaptable up to a certain point and is not sufficient for dealing with evolving cyber threats.

## Risk Quantification and Assessment Framework

Overall Impact on Security - **Low**

Creating a national risk assessment framework would allow utilities to identify and evaluate risks to their security. This would enable utilities to identify and prioritize vulnerabilities systematically, allowing for more effective defense resource usage and risk mitigation. Risk assessment does not directly prevent attacks, so its impact on overall security is somewhat limited compared to other policies [10].

Implementation Cost - **Medium**

Developing and implementing a risk quantification framework would require government investments in cybersecurity expertise, data analytics tools, and staff training. Utilities would also need to acquire the necessary tools and knowledge to assess and manage risks systematically. More efficient risk mitigation would offset Some of these costs, but this policy would have a considerable cost to both the government and utilities [22].

Feasibility - **Medium**

Utilities operate with different infrastructure technologies, making it challenging to create a system that can accurately assess risks throughout the entire grid. In addition, larger utilities with more resources are more likely to use a risk quantification framework, while smaller utilities that can't afford the tools necessary for the program may be left unable to participate. Creating a framework that can be used by small and large utilities while remaining useful is an issue for this policy [22].

Adaptability - **Medium**

This policy is capable of evolving with new threats and vulnerabilities. A risk quantification framework can be updated to account for trends in cyberattacks and can be customized for the particular application in which it is being used. This adaptability is a strength of the policy. However, this relies on proper implementation, maintenance, and support from the government [23].

## Cyber Disaster Recovery Plan

Overall Impact on Security - **Medium**

Creating a cyber disaster recovery plan would enhance the power grid's ability to respond to and recover from cyberattacks by incorporating incident response protocols, backup and restoration procedures, redundancy and failover systems, clear communication plans, and regular drills. Many utilities currently rely on ad hoc responses to cyber incidents, leaving them vulnerable to prolonged downtime and service disruption. A recovery plan would minimize service interruptions, ensure faster power restoration, and maintain public trust in the grid's reliability. A cyberattack recovery plan is impactful but does not inherently affect security [24].

Implementation Cost - **High**

Implementing a cyber disaster recovery plan would involve significant costs, including backup systems, disaster response training, and testing protocols. Heavy federal funding or subsidies would be necessary to support the development and maintenance of the policy. There is potential for billions in savings if the recovery plan can help alleviate a crisis, but the maintenance would be expensive regardless [25].

Feasibility - **Medium**

Larger utilities with access to more resources would be able to implement robust disaster recovery plans, but smaller utilities would need significant support. This is true, especially when replacing particular components, such as transformers, which are expensive and difficult to acquire. Successful implementation would depend on extensive coordination and phased rollout across the grid [24].

Adaptability - **High**

A cyber disaster recovery plan's adaptability would depend on how often the plan is updated and tested to reflect new threats and evolving technologies. If properly implemented, it would be highly adaptable and capable of addressing a variety of cyberattack scenarios. Since this policy focuses on recovery rather than prevention, it should be implemented with other proactive cybersecurity measures to provide comprehensive protection.

## Enhanced Information Sharing

Overall Impact on Security - **High**

The government currently has information-sharing mechanisms, but their value is hindered by a lack of trust and collaboration between utilities, government agencies, and third-party stakeholders. Enhancing information sharing programs with secure communication methods and legal protection to address liability concerns would encourage more utilities to participate, providing them with knowledge of emerging threats. This timely and secure threat intelligence sharing would significantly strengthen collective defense, leading to more effective cybersecurity across the grid [26].

Implementation Cost - **Medium**

The cost of this policy is from building secure information-sharing platforms and overcoming legal barriers. The government would need to develop a centralized, secure platform for sharing threat intelligence. This would require significant investment in secure communication technologies and protocols to protect sensitive data. Additionally, the government would need to establish clear guidelines for data sharing, including privacy protections and liability limitations for utilities participating in the program [26].

Feasibility - **Medium**

The feasibility of this policy is medium due to the legal and trust barriers between utilities and government agencies. The government would need to pass federal legislation encouraging information sharing while providing legal protections for utilities, ensuring that companies do not fear penalties or for sharing vulnerabilities in the program. Addressing these concerns through federal legislation and trust-building measures would be critical for success [27].

Adaptability - **High**

This policy has strong adaptability potential because information-sharing platforms are designed to communicate evolving threats and technological advancements. Continuous system updates and real-time threat intelligence sharing would ensure that utilities stay informed and respond quickly to new types of attacks. A flexible platform would enhance resilience, making the grid better equipped to handle future cyber threats [27].

## Secure the Supply Chain

Overall Impact on Security - **High**

Securing the supply chain for critical power grid components would address a significant vulnerability in national grid security. By implementing regulations that require sourcing from trusted vendors, conducting security audits, and enhancing hardware tracking, this policy would mitigate the risk of supply chain-based cyberattacks. This is crucial for minimizing threats from foreign vendors that may attempt to introduce malicious components into the grid. Ensuring that only secure, vetted components are used would increase the overall security of the U.S. power grid [28]. The DOE and the Federal Energy Regulatory Commission would be responsible for developing and enforcing these regulations, while utilities and vendors would be tasked with compliance.

Implementation Cost - **High**

The cost of securing the supply chain would be substantial. The power grid relies on complex components like transformers and SCADA systems, which are difficult to source securely because many come from international suppliers. Additionally, vetting each supplier to ensure cybersecurity compliance is resource-intensive, requiring background checks, risk assessments, and monitoring. These processes require financial investment and skilled personnel, driving up the costs of securing the power grid's supply chain [29].

Feasibility - **Low**

The low feasibility of this policy is due to the complex coordination needed between domestic and international vendors and the potential political challenges that arise from restricting certain suppliers. International vendors would need to comply with U.S. security standards, and federal agencies would have to establish clear enforcement mechanisms to ensure adherence. Attempting to use these enforcement mechanisms would complicate international trade and relations. While important, the difficulty of enforcing this policy lowers its feasibility [28].

Adaptability - **Medium**

Securing the supply chain is somewhat adaptable due to its focus on regulations and vendor management. As threats evolve and new vulnerabilities emerge, security standards and supplier vetting processes can be updated to address these changes. The ability to track and audit hardware would also help utilities adapt to emerging cyber threats. However, the policy is limited by the complexity of coordinating international compliance.

## Increased Funding and Support for Smaller Utilities

Overall Impact on Security - **Medium**

Small utilities have numerous cybersecurity vulnerabilities due to a lack of funding and trained staff. Many of these utilities welcome security standards and guidelines but fall short of implementation for financial reasons. This policy would help address this issue by providing more resources to small utilities, bringing them up to par with larger utilities [30]. While supporting small utilities is important, it does not offer as much overall security to the power grid as other policies.

Implementation Cost - **Medium**

The main expenses involve providing funds to hire cybersecurity professionals, upgrade outdated systems, and provide training to staff. An issue with this policy is that there is no set cost. Small utilities will never refuse more government funds and will be reluctant to give up extra money once they reach cybersecurity standards. Support and maintenance costs will also require ongoing funding. This policy has the potential to become extremely costly if not managed carefully [31].

Feasibility - **Medium**

While the distribution of government resources is straightforward, successful implementation requires ensuring that smaller utilities effectively use these funds. The government can use grants instead of dollars and attempt to ensure compliance with audits and performance reporting, but there is no guarantee that money will be used appropriately. Smaller utilities may also need more technical expertise or administrative capacity to meet the conditions tied to public funding, complicating the overall rollout and reducing feasibility [31].

Adaptability - **Low**

This policy has low adaptability because it focuses on smaller utilities, which won't significantly enhance overall grid security compared to other policies. Additionally, managing compliance and ensuring effective use of funds adds complexity, requiring extensive oversight and administration, which limits the policy's flexibility. These factors make it difficult to scale or adjust the policy to address future threats.

# Policy Recommendation

## Policy Matrix Methodology

Policies have been listed across the X axis with the criteria used to rate them along the Y axis. The columns below each policy have been filled with the assessed scores from above. Ratings have been assigned a numerical value (High = 3, Medium = 2, and Low = 1) so that we can find an average score for each policy. This average score can be seen in the "Overall Rating" row. Policies with a higher overall rating are more effective in bolstering power grid cybersecurity compared to policies with a lower overall rating. Note that implementation cost is weighed inversely (High = 1, Medium = 2, and Low = 3).

## Policy Matrix

| Criteria | Unified Framework | Cyber Hygiene | Risk Quantification | Recovery Plan | Information Sharing | Secure Supply Chain | Support Small Utilities |
|---|---|---|---|---|---|---|---|
| Overall Impact on Security | High | Medium | Low | Medium | High | High | Medium |
| Implementation Cost | High | Low | Medium | High | Medium | High | Medium |
| Feasibility | Medium | High | Medium | Medium | Medium | Low | Medium |
| Adaptability | High | Medium | High | High | High | Medium | Low |
| | | | | | | | |
| | | | | | | | |
| Overall Rating | 2.25 | 2.5 | 2 | 2 | 2.5 | 1.75 | 1.75 |

## Overall Recommendation

Based on the policy matrix, the federal government should prioritize the implementation of three key policies to enhance the cybersecurity posture of the U.S. power grid:

The government must mandate baseline cyber hygiene practices, including strong password requirements, phishing response, social engineering training, regular system updates, and basic network segmentation. This could be accomplished efficiently by NIST, which already provides a cybersecurity framework that can be modified to provide baseline standards. NERC would ensure these standards are followed through audits and non-compliance penalties. These cost-effective measures are a quick fix for common vulnerabilities exploited across the power grid.

In addition, the government must create a secure, real-time information-sharing platform between utilities, government agencies, and stakeholders to improve threat detection and response. It should feature secure communication protocols and legal protections to address privacy and liability concerns. The Electricity Information Sharing and Analysis Center can lead this initiative by updating its current infrastructure to make collaboration more attractive to utilities. Enhanced information sharing will strengthen collective defense, improve response to cyberattacks, and build trust among utilities and the government.

Finally, to ensure robust cybersecurity for the future, the U.S. must establish a unified national cybersecurity framework to eliminate regulatory overlaps and ensure consistent security protocols. This framework will standardize best practices and provide clear guidance for utilities on how to implement cybersecurity measures. To further ensure effectiveness, this framework could include provisions for reassessments and updates, requiring utilities to remain vigilant and adapt to evolving cyber threats. By establishing consistent protocols, the U.S. can proactively defend its power grid, making it more resilient to both current and future cyberattacks.

# Conclusion

Safeguarding the U.S. power grid from cyber threats is a complex challenge that requires both immediate and sustained policy action. The current fragmented cybersecurity standards expose critical infrastructure to significant vulnerabilities. By implementing a dual policy approach focused on enhanced information sharing and standardized cyber hygiene practices, the U.S. can mitigate common vulnerabilities and build collective defenses among utilities. This will make significant progress toward improving grid security in the short term.

To ensure long-term security, establishing a Unified National Cybersecurity Framework is crucial. By consolidating existing standards, this framework will streamline defense, close security gaps, and ensure that all utilities, regardless of size or location, adhere to a cohesive set of protocols. This baseline will not only safeguard critical infrastructure today but also enable America to strengthen its defenses in the future, securing national security, public safety, and economic stability.

# Sources